\documentclass[aps,prl,twocolumn,showpacs,superscriptaddress,groupedaddress]{revtex4}  
\usepackage{graphicx}  
\usepackage{dcolumn}   
\usepackage{bm}        
\usepackage{amssymb}   
\usepackage{amsmath}   

\begin{document}



\title{Dark energy in thermal equilibrium with the cosmological horizon?}

\author{Vincent Poitras}
\address{McGill University,\\3600 University street, Montreal, Canada}

\date{\today}

\begin{abstract}
According to a generalization of black hole thermodynamics to a cosmological framework, it is possible to define a temperature for the cosmological horizon. The hypothesis of thermal equilibrium between  the dark energy and the horizon has been considered by many authors. We find the restrictions imposed by this hypothesis on the energy transfer rate ($Q_i$) between the cosmological fluids, assuming that the temperature of the horizon have the form $T=b/2\pi R$, where $R$ is the radius of the horizon. We more specifically consider two types of dark energy: holographic dark energy (HDE) and dark energy with a constant EoS parameter ($w$DE). In each case, we show that for a given radius $R$, there is an unique term $Q_{de}$ that is consistent with  thermal equilibrium. We also consider the situation where, in addition to dark energy, other fluids (cold matter, radiation) are  in thermal equilibrium with the horizon. We find that the interaction terms  required for this will generally violate  energy conservation ($\sum_i Q_i=0$).
\end{abstract}


\maketitle

\section{I. Introduction}
\label{sect:Intro}
In the late 1990s, observations of supernovae \cite{SNE1,SNE2,SNE3} have suggested that the Universe is undergoing a state of accelerated expansion. Since then, additional evidence leading to the same conclusion have been found \cite{SNE4,SNE5,CMB1,CMB2,CMB3,CMB4,planck,BAO1,BAO2}. In the context of general relativity (GR), the equation of state (EoS) parameter of a fluid, defined as the ratio of its pressure over its energy density ($w\equiv p/\rho$), must be smaller than $-1/3$ in order to be able to drive the accelerated expansion of the Universe. Since normal matter satisfies the strong energy condition ($w\geq0$), this condition is not fulfilled. Hence, two main approaches have been proposed to explain the acceleration: one of them consists to replace the GR by a modified gravity theory (see e.g. Refs.~\cite{ModG1,ModG2,ModG3}) and the other, to keep GR while introducing a new cosmic fluid, known as dark energy (see Ref.~\cite{WYI} and references therein), endowed with a sufficiently large and negative EoS parameter (${w_{de}<-1/3}$). Alternatively, it has also been proposed that the (apparent) acceleration could be only an artifact caused by the spatial inhomogeneity of the Universe (see e.g. Refs.~\cite{Inhomo1,Inhomo2}). 

The $\Lambda$CDM model is the simplest cosmological model which provides a reasonably good fit to the observational data. In this model, the two main components of the Universe are currently a form of dark energy provided by a cosmological constant ($\Lambda$) and a  pressureless fluid known as cold dark matter (CDM). In addition to these two fluids, the Universe is also composed of ordinary matter (radiation, baryons). However, despite the excellent agreement with the observational data, the $\Lambda$CDM model is facing two theoretical difficulties. The most serious one concerns the value of the dark energy density and is known as the \emph{cosmological constant problem} \cite{Weinberg}. Indeed, there is a discrepancy of $\sim$123  orders of magnitude between the value expected from theoric computations and the value inferred from observations ($\rho_{de_{\rm{obs}}}/\rho_{de_{\rm{th}}}\sim10^{-123}$). The other one, dubbed  \emph{coincidence problem} \cite{CoinProb}, relies on the observation that the values of the matter energy density and of the dark energy density are currently of the same order of magnitude. Unlike to the previous problem, this  is not incompatible with the theory. However, since the matter energy density is diluted proportionally to the volume of the Universe as it is expanding ($\rho_{m}\propto a^{-3}$) while the dark energy density remains constant ($\rho_{de}=const$), the period of time during which $\rho_{m}/\rho_{de}=\mathcal{O}(1)$ corresponds to a very narrow window in the Universe history. To currently lie in this window, a fine tuning of the initial conditions of the model is needed. However, it is worth mentioning that about a decade before the discovery of the accelerated expansion of the Universe, anthropic arguments were already addressing both problems \cite{Wein0,Weinberg}.

A possible way to circumvent these problems, without the recourse to anthropic arguments, would be to allow the dark energy density to vary in time ($\rho_{de}\neq const$). It would then be possible for the dark energy density to decrease from an initial large value, consistent with the theoritical computation, to a smaller one, consistent with the current value inferred from the observations. Moreover, that could also extend duration of the period  during which $\rho_{m}/\rho_{de}=\mathcal{O}(1)$. A variable dark energy density could be obtained or by considering an EoS parameter $w_{de}$ different from $-1$, either by allowing an energy transfer between the dark energy and another fluid (of course, these two ways could be considered together). Several forms of dark energy models have been proposed, including  quintessence \cite{Quint1}, phantom fields \cite{Phant1}, tachyon fields \cite{Tach1}, Chaplygin gas \cite{ChapG1}, agegraphic dark energy \cite{AgeDE1} and holographic dark energy \cite{UV1}, to name few.

To study the thermodynamical implications of these models, the determination of the dark energy temperature  is a question that must inevitably be addressed. A hypothesis often used \cite{Tprop02,Tprop03}  is that the dark energy temperature is proportional to that of the cosmological horizon ($T_{de}\propto T_{h}$). Indeed, according to a generalization of black-hole thermodynamics to a cosmological framework, it is possible to define a temperature for the horizon which is related to its surface gravity (see section \ref{sect:cht} for more details). A stronger hypothesis \cite{ThEq0,ThEq1,ThEq2,ThEq3,ThEq4,ThEq5,ThEq6,ThEq7,ThEq8,FF01,FF02,FF03,N02}, albeit more motivated, consists in considering that the dark energy fluid and the horizon are in thermal equilibrium ($T_{de}=T_{h}$). An argument presented in Ref.~\cite{ThEq0} and reused in Refs.~\cite{ThEq1,ThEq2,ThEq3,ThEq4,ThEq5,ThEq6,ThEq7,ThEq8} states that if this were not the case, then the ``energy would spontaneously flow between the horizon and the fluid (or vice versa), something at variance with the FRW geometry''. Following this argument, some authors ~\cite{ThEq5,ThEq6,ThEq7,ThEq8}  have even extended this hypothesis to the other fluids, assuming that the thermal equilibrium between the horizon and a given fluid must hold at least for late time. 

Although this assumption may be questionable (especially in regard to its extension to other fluids), the objective of this paper is not to directly discuss of its validity. Instead, we will demonstrate that in order to maintain  thermal equilibrium between a given fluid and the horizon, a  specific energy transfer rate is required, which constitutes a highly restrictive condition for its application.

\section{II. Dynamics}
\subsection{A. Interacting fluids}
\label{sect:IDS}

In a Friedmann-Robertson-Walker (FRW) spacetime, the continuity equations for a model allowing interactions between the different cosmic fluids (dark energy, dark matter, baryonic matter and radiation)  are given by
  \begin{equation}
     \dot{\rho}_i + 3H(1+w_i)\rho_i= Q_{i}.
     \label{eq:cont}
  \end{equation}
If we treat the curvature as fictitious fluid, this equation can also be used to describe the evolution of its energy density $\rho_k\equiv-3k/8\pi G a^2$ \footnotetext[1]{The curvature parameter $k$, whose dimensions are (length)$^{-2}$, is negative for an open Universe and positive for a closed one.}$^1$. Since the Hubble term is defined as $H=\dot{a}/a$, where $a$ is the scale factor, in absence of interaction ($Q_i=0$), the solution to this equation is
  \begin{equation}
     \rho_i=\rho_{i_0}a^{-3(1+w_i)}.
     \label{eq:rho}
  \end{equation}
Here, we have set $a_0=1$ (in this paper, the subscript 0 refers to the current value of a variable). The Friedmann equations can be written as
    \begin{equation}
      H^2=\dfrac{M_p^{-2}}{3}\sum_{i} \rho_{i},
     \label{eq:fried1}
  \end{equation}
   \begin{equation}
      \dot{H}=-\dfrac{M_p^{-2}}{2}\sum_i(1+w_i) \rho_i,
     \label{eq:fried2}
  \end{equation}
where $M_p=(8\pi G)^{-1/2}$ is the reduced Planck mass (throughout this work, we will use a unit system where $\hbar=k_B=c=1$). The LHS of Eq.~(\ref{eq:cont}) has the same form as in the  non-interacting case, where $H\equiv\dot{a}/a$ ($a$ is the scale factor) stands for the Hubble term, $\rho_i$, for the energy density of a given fluid and $w_i\equiv p_i/\rho_i$ ($p_i$ is the pressure), for the equation of state (EoS) parameter of this fluid. The values of these parameters are the  usual ones for radiation  ($w_{r}=1/3$), for curvature (${w_k=-1/3}$) and, in absence of interaction (see section~\ref{sec:ofluid}), for dark and baryonic matter ($w_{dm}=w_{b}=0$). For dark energy, $w_{de}$ is not necessarily fixed to $-1$ as in the $\Lambda$CDM model and could even be variable. The RHS of the equation represents the possible interactions between the fluids. A positive value ($Q_i>0$) represents a gain of energy for the fluid, and negative value ($Q_i<0$), a loss. The ensemble of these terms is subject to the energy conservation condition $\sum_{i}Q_{i}=0$. It is to be noticed that the interaction is allowed only between the \emph{real} fluids. For the curvature, the interaction term $Q_k$  must be zero, otherwise it would imply that the curvature parameter $k$ is variable, which would be inconsistent with the FRW metric.

\subsection{B. Types of dark energy}
The exact nature of dark energy is not known, so it is not uncommon to see it studied from a purely phenomenological point of view. Among the ideas which could help us to study dark energy from a more fundamental point of view, there is the holographic principle. This  states that the number of degrees of freedom of a physical system  must scale with the area of its  boundary \cite{holop1,holop2}. This principle also suggests  a connection between the UV cutoff scale ($\Lambda_{UV}$) of an effective quantum field theory and an IR length corresponding to the size of a system ($R$). Indeed, Cohen et al. \cite{Cohen} have argued that the zero-point energy of a system  should not exceed the mass of a same-size black hole
  \begin{equation}
      R^3\Lambda_{UV}^4\lesssim RM^2_P.
     \label{eq:ineq}
  \end{equation}
In a cosmological context,  the energy density $\Lambda_{UV}^4$ has been identified with the dark energy density $\rho_{de}$ \cite{UV1,UV2,UV3}. The largest $R$ allowed is the one saturating this inequality, so we can write
  \begin{equation}
      \rho_{de}=\rho_{de_0}\left(\dfrac{R_0}{R}\right)^2,
     \label{eq:hde}
  \end{equation}
where $\rho_{de_0}=3\alpha^2M_p^2/R_0^2$. The dimensionless factor $\alpha$ is usually assumed constant (for some examples where it is not the case, see \cite{varb1} and references therein) and this type of dark energy is known as holographic dark energy (HDE). Corrections  based on entropic considerations have been discussed in the literature \cite{ECHDE}, but in this paper, we will only consider the form of HDE given by Eq.~(\ref{eq:hde}). How to define the size of the system in a cosmological context  is not a simple question and different choices has been discussed in the literature. The most common are the Hubble radius (${R_H=1/|H|}$) \cite{UV3,RH1}, the apparent horizon radius  (${R_A=1/\sqrt{H^2-\frac{1}{3}M_p^{-2}\rho_k}}$) \cite{ThEq6}, the event horizon radius (${R_E=a\int_{t}^{t_{end}}\frac{dt}{a}}$)\footnotetext[2]{The upper integration limit is given by $t_{end}=\infty$ in an eternally expanding model and by  the time of the big crunch in a recollapsing model. This  expression may also be computed as ${R_E=a\int_{a}^{a_{end}}\frac{da}{H^2a}}$, where $a_{end}=a(t_{end})$.}$^{2}$ \cite{UV1} and the Ricci length (${R_{CC}=1/\sqrt{\dot{H}+2H^2}}$) \cite{RC1} (and its modified definition \cite{MRC1}). In this paper, unless otherwise indicated, we will consider a generic length $R$.
 
Using Eqs.~(\ref{eq:cont}) and (\ref{eq:hde}), it is straightforward to show that the EoS parameter of HDE is given by
  \begin{equation}
      w_{de}=-1 + \dfrac{1}{3HR}\left[2\dot{R}-\frac{Q}{3\alpha^2M_p^2}R^3\right].
     \label{eq:wde}
  \end{equation}
This expression will generally be  variable. Hence, as a complement to HDE, we will consider a second type of dark energy  for which the EoS parameter has a fixed value ($w$DE).

\section{III. Thermodynamics}

\subsection{A. Cosmological horizon temperature}
   \label{sect:cht}

Since the seminal works of Hawking \cite{BH1} and Bekenstein \cite{BH2} in the seventies, the thermodynamical properties of black holes have been widely studied. One of the most well known feature is that, as consequence of the existence of an event horizon, the stationary (or quasi-stationary) black holes behave like black bodies emitting  thermal radiation with a temperature proportional to the value of the surface gravity evaluated on the horizon 
  \begin{equation}
      T_h=\dfrac{\kappa}{2\pi}.
     \label{eq:kh}
  \end{equation}
A first extension of black hole thermodynamics to a cosmological framework was done by Gibbons and Hawking in Ref.~\cite{BH3} by considering  de Sitter space. In this case, the surface gravity on the event horizon is given by the inverse of the horizon radius, $\kappa=1/R_E=\sqrt{3/\Lambda}$, thus the temperature is given by
  \begin{equation}
      T_h=\dfrac{1}{2\pi R_E}.
     \label{eq:TE}
  \end{equation}
Unlike to  de Sitter space, the event horizon is not always well defined for FRW spacetime. However, it has been argued \cite{ArgRA1,ArgRA2} that it is actually the apparent horizon, and not the event horizon, that is responsible for Hawking radiation (in the case of de Sitter space, the two horizons coincide). It worth mentioning that for  de Sitter space,  the event horizon radius  has a constant value, while for a FRW spacetime, the value of the apparent horizon radius varies. To compute the surface gravity, this could be problematic.  Indeed, this quantity is usually defined in terms of Killing horizons, which work well in stationary (or quasi-stationary) situations. For the dynamical situations where no such horizons exist, several  definitions have been proposed (see \cite{SurfGr01,SurfGr03} for a review). If we consider a generic spherically symmetric spacetime, the line element is given by
  \begin{equation}
      ds^2=h_{ab}dx^adx^b+\tilde{r}^2d\Omega^2,
     \label{eq:FRWm2}
  \end{equation}
where $x^0=t$, $x^1=r$, $\tilde{r}=a(t)r$ and $d\Omega^2=d\theta^2 + \sin^2{\theta}d\phi^2$. For the FRW spacetime, the 2-dimensional metric $h_{ab}$ is given by ${\rm{diag}}(-1,a^2/(1-kr^2))$. A frequently used definition of the surface gravity has been proposed by Hayward in Ref.~\cite{SurfGr02}:                                 
  \begin{equation}
      \kappa=\dfrac{1}{2}\nabla\cdot\nabla\tilde{r}=\dfrac{1}{2\sqrt{-h}}\partial_a\left(\sqrt{-h}h^{ab}\partial_b\tilde{r} \right).
     \label{eq:dynsurfg}
  \end{equation}
Here, the divergence and gradient refer to the two-dimensional space normal to the spheres of symmetry. An evaluation of this expression at $\tilde{r}=R_A$ gives $\kappa=(1-\epsilon)/R_A$, where $\epsilon\equiv\dot{R}_A/(2HR_A)$. Thus the horizon temperature is given by
  \begin{equation}
      T_h=\dfrac{1-\epsilon}{2\pi R_A}.
     \label{eq:TAeps}
  \end{equation}
An alternative definition \cite{Tprop03} for the dynamical surface gravity is 
  \begin{equation}
     \kappa=-\dfrac{1}{2}\partial_{\tilde{r}}\chi=\dfrac{\tilde{r}}{R_A^2},
     \label{eq:dynsurfg2}
  \end{equation}
where $\chi\equiv h^{ab}\partial_a{\tilde{r}}\partial_b{\tilde{r}}$ \footnotetext[3]{It is to be noticed that the radius of the apparent horizon, $R_A$, is defined as the value of  $\tilde{r}$ for which the scalar $\chi$ vanishes (which implies that the vector $\nabla\tilde{r}$ is null on the apparent horizon surface).}$^3$. At $\tilde{r}=R_A$, the surface gravity is then given by $\kappa=1/R_A$, and the horizon temperature by
  \begin{equation}
      T_h=\dfrac{1}{2\pi R_A}.
     \label{eq:TA}
  \end{equation}
Among the papers where a thermal equilibrium between the horizon and the dark energy is considered, both Eq.~(\ref{eq:TAeps}) \cite{ThEq6,ThEq8,FF01,FF02} and Eq.~(\ref{eq:TA}) \cite{ThEq1,ThEq3,ThEq4,ThEq7,FF03}  are commonly used as a definitions of the horizon temperature. Although it has been argued \cite{Cai2005} that the $\epsilon$ term can be neglected in certain situations, these two expressions are generally different and one can wonder whether  one definition is better motivated than the other. In favor of Eq.~(\ref{eq:TA}), it was shown in Ref.~\cite{Cai2009}, using the tunneling approach, that an observer inside the apparent horizon of a FRW Universe will see a thermal spectrum with a temperature given by ${T_h=1/(2\pi R_A)}$, without the extra $\epsilon$ term. It is also interesting to notice that using this expression for the temperature, it is possible to recover the second Friedmann equation (Eq.~(\ref{eq:fried2})) from the first law of thermodynamics \cite{Cai2005}. Some authors still  consider the event horizon as the relevant one and use Eq.~(\ref{eq:TE}) to define the horizon temperature \cite{ThEq0,ThEq2,ThEq5} (see however Ref.~\cite{N02} where Eq.~(\ref{eq:dynsurfg2}) is evaluated at $\tilde{r}=R_E$, which leads to ${T_h=R_E/(2\pi R_A^2)}$). In Refs.~\cite{ThEq4,FF03,Tprop03}, the horizon temperature is assumed to be proportional to its de Sitter value, i.e.
  \begin{equation}
      T_h=\dfrac{b}{2\pi R_H},
     \label{eq:TH}
  \end{equation}
where $b$ is a constant parameter. It would be interesting to consider all these different definitions, but for the sake of conciseness we will restrict our attention (while keeping in mind that there is  no clear consensus on how the horizon temperature should be defined and which horizon should be considered) to the case where the temperature has the dependence on the horizon radius given by
  \begin{equation}
      T_h=\dfrac{b}{2\pi R}.
     \label{eq:Thor}
  \end{equation}
Here $R$ could stand for, with $b=1$, the event horizon radius (Eq.~(\ref{eq:TE})) and the apparent horizon radius (Eq.~(\ref{eq:TA})), as well for  the Hubble radius (Eq.~(\ref{eq:TH})).

\subsection{B. Conditions for thermal equilibrium}
 \label{sec:cond}
To find the form of the energy transfer rate $Q_i$ required to maintain thermal equilibrium between a fluid, whose the continuity equation is given by  Eq.~(\ref{eq:cont}), and the cosmological horizon, we will first derive an equation for the temperature evolution for this fluid. Our derivation is similar to that presented in Ref.~\cite{Maart}. The starting point is the Gibbs equation, $T_idS_i=dE_i+p_idV$. For simplicity we will consider a comoving volume of unit coordinate volume and hence a physical volume of $V=a^3$. Since the energy of the fluid is given by $E_i=\rho_iV$, we can rearrange the Gibbs equation as
  \begin{equation}
    dS_i=\dfrac{\rho_i+p_i}{T_i}dV+\dfrac{V}{T_i}d\rho_i.
  \label{eq:Gibbs}
  \end{equation}
From this expression for the entropy, we can show that the integrability condition
  \begin{equation}
    \left[\dfrac{\partial }{\partial V}\left(\dfrac{\partial S_i}{\partial T_i}\right)_{N_i,V}\right]_{N_i,T_i}=\left[\dfrac{\partial }{\partial T_i}\left(\dfrac{\partial S_i}{\partial V}\right)_{N_i,T_i}\right]_{N_i,V}
  \label{eq:intcond1}
  \end{equation}
implies that
  \begin{equation}
    T_i\left(\dfrac{\partial p_i}{\partial T_i}\right)_{N_i,V}=(\rho_i+p_i)+V\left(\dfrac{\partial \rho_i}{\partial V}\right)_{N_i,T_i}.
  \label{eq:intcond1.5}
  \end{equation}
Except for the cases where the derivatives vanish or are ill-defined (e.g. for the DE in $\Lambda$CDM model) this equation is equivalent to
  \begin{equation}
    T_i\left(\dfrac{\partial p_i}{\partial \rho_i}\right)_{N_i,V}=(\rho_i+p_i)\left(\dfrac{\partial T_i}{\partial \rho_i}\right)_{N_i,V}-V\left(\dfrac{\partial T_i}{\partial V}\right)_{N_i,\rho_i}.
  \label{eq:intcond2}
  \end{equation}
Since we can express the temperature as a function of the volume and the energy density ($T_i=T_i(\rho_i,V)$), its time derivative  may be expressed as ${\dot{T}_i=(\partial{T_i}/\partial{\rho_i})\dot{\rho_i}+(\partial{T_i}/\partial{V})\dot{V}}$. The time derivative of the physical volume $V=a^3$ is $\dot{V}=3HV$; then using also Eq.~(\ref{eq:cont}) to replace $\dot{\rho_i}$,  we get
  \begin{equation}
  \begin{split}
    \dot{T}_i=-3H\left[(\rho_i+p_i)\left(\dfrac{\partial T_i}{\partial \rho_i}\right)_{N_i,V}-V\left(\dfrac{\partial T_i}{\partial V}\right)_{N_i,\rho_i}\right]\\
     + Q_i\left(\dfrac{\partial T_i}{\partial \rho_i}\right)_{N_i,V}
  \label{eq:dTi1}
  \end{split}
  \end{equation}
The expression in the square brackets is identical to the RHS of Eq.~(\ref{eq:intcond2}), hence we can write the temperature evolution equation as
  \begin{equation}
    \dfrac{\dot{T}_i}{T_i}=-3H\left(\dfrac{\partial p_i}{\partial \rho_i}\right)_{N_i,V} + \dfrac{Q_i}{T_i}\left(\dfrac{\partial T_i}{\partial \rho_i}\right)_{N_i,V}.
  \label{eq:dTi2}
  \end{equation}
Now to find the form of the energy transfer rate required to have  thermal equilibrium ($\tilde{Q}_i$) between the cosmic fluid and the cosmological horizon ($T_i=T_h\equiv T)$, we must simply solve the preceding equation for $Q_i$ and replace the temperature by the expression given by Eq.~(\ref{eq:Thor}),
  \begin{equation}
    \tilde{Q}_i=\dfrac{b}{2\pi}\left(\dfrac{\partial\rho_i}{\partial T}\right)_{N_i,V}\left[\dfrac{3HR(\partial p_i/\partial\rho_i)_{N_i,V}-\dot{R}}{R^2}\right].
  \label{eq:QthEq}
  \end{equation}
In the following sections, we will evaluate this expression for the two different types of dark energy (HDE and $w$DE), as well for relativistic and non-relativistic matter.

\subsubsection{1. Holographic dark energy}
If we assume that the HDE is in thermal equilibrium with the horizon ($T_{hde}=T_h$), it is straightforward to show from Eqs.~(\ref{eq:hde}) and (\ref{eq:Thor}) that the energy density of HDE depends only on its temperature
  \begin{equation}
\rho_{hde}=\rho_{hde_0} \left(\dfrac{T_{hde}}{T_{hde_0}}\right)^2.
     \label{eq:rhdeT}
  \end{equation}
where $T_{hde_0}=b/(2\pi R_0)$. Here following Refs.~\cite{ThEq2,ThEq5}, we have assumed that the radius involved in the definitions of the HDE and  of the horizon temperature are the same. 
To find an expression for the HDE pressure, we can invert this equation and insert the expression thus obtained for the temperature in Eq.~(\ref{eq:intcond2}), which becomes
  \begin{equation}
     \dfrac{d p_{hde}}{d\rho_{hde}}=\dfrac{w_{hde}+1}{2}.
      \label{eq:diffp}
  \end{equation}
Here,  we have replaced the partial derivative by a total derivative. Since the HDE energy density is only a function of the temperature, it seems reasonable to assume that it is also the case for the HDE pressure. If we consider a constant EoS parameter, the only possible solution to the above equation is $w_{hde}=1$. In this case, the HDE and  the $w$DE are equivalent. However, with this value, the Universe will never experience a phase of accelerated expansion and we must therefore reject it. If we consider instead a variable EoS parameter,  Eq.~(\ref{eq:diffp}) can be more conveniently written as
  \begin{equation}
     \dfrac{d w_{hde}}{d\rho_{hde}}=\dfrac{1-w_{hde}}{2\rho_{hde}}
      \label{eq:diffp2}
  \end{equation}
and its solution is given by
  \begin{equation}
     w_{hde}=1+(w_{hde_0}-1)\left(\dfrac{\rho_{hde_0}}{\rho_{hde}}\right)^{\frac{1}{2}}. 
     \label{eq:whde}
  \end{equation}
The HDE pressure is then simply given by $p_{hde}=w_{hde}\rho_{hde}$. Having found an expression for the pressure and for the energy density, we come back to Eq.~(\ref{eq:QthEq}), which becomes for the HDE 
  \begin{equation}
      \tilde{Q}_{hde}=\left[\dfrac{3(1+w_{hde})HR-2\dot{R}}{R} \right]\rho_{hde}.
     \label{eq:Qhde}
  \end{equation}
We note that we would have obtained the exact same expression for $Q$ by inserting the HDE energy density (Eq.~(\ref{eq:hde})) in the  continuity equation~(\ref{eq:cont}). Therefore,  the expression for the interaction term given by Eq.~(\ref{eq:Qhde}) is always true for HDE; however that will imply a thermal equilibrium between the HDE and the cosmological horizon only if the EoS parameter has the form given by Eq.~(\ref{eq:whde}).

\subsubsection{2. Dark energy with a constant EoS parameter}
\label{sect:wDE}
Like for the HDE, we will assume that the energy density (and  consequently the pressure) of the $w$DE depends only on the temperature. In this case, we can replace the first partial derivative in Eq.~(\ref{eq:QthEq}) by a total derivative and write $d\rho_{wde}/dT=\dot{\rho}_{wde}/\dot{T}$. This leads, after some simple manipulations, to 
 \begin{equation}
      \tilde{Q}_{wde}=\dot{\rho}_{wde}\left[1-3w_{wde}H\dfrac{R}{\dot{R}} \right]. 
     \label{eq:Qde1}
  \end{equation}
Using the continuity  equation~(\ref{eq:cont}) to replace $Q_{wde}$, we obtain (for $w_{wde}\neq-1$) the  differential equation 
  \begin{equation}
     \dfrac{\dot{\rho}_{wde}}{\rho_{wde}}=-\left(\dfrac{1+w_{wde}}{w_{wde}}\right)\dfrac{\dot{R}}{R},
     \label{eq:diffeq}
  \end{equation}
whose solution is
  \begin{equation}
      \rho_{wde}=\rho_{de_0}\left(\dfrac{R}{R_{0}}\right)^{-\frac{1+w_{wde}}{w_{wde}}}.
    \label{eq:rwde}
  \end{equation}
Inserting this expression into Eq.~(\ref{eq:Qde1}) yields
  \begin{equation}
      \tilde{Q}_{wde}=\left[\dfrac{3(1+w_{wde})HR-(1+w_{wde}^{-1})\dot{R}}{R} \right]\rho_{wde}.
     \label{eq:Qwde}
  \end{equation}
This expression is valid for any constant EoS parameter except $w_{wde}=-1$. We recover the dark energy of the $\Lambda$CDM model for this value ($\rho_{wde}=const$ and $Q_{wde}=0$), but we cannot conclude that thermal equilibrium with the horizon is possible for this type of dark energy since, as  was pointed in section~\ref{sec:cond}, our derivation is not valid for a fluid whose  energy density and  pressure are intrinsically constant (in this case, we can even ask whether a temperature can be meaningfully defined).

\subsubsection{3. Other fluids}
 \label{sec:ofluid}

As mentioned above, some authors ~\cite{ThEq5,ThEq6,ThEq7,ThEq8,FF01,FF02,FF03} considered the possibility that, in addition to dark energy, other fluids could also be in thermal equilibrium with the horizon. We will now consider the implications of this hypothesis. For an ultra-relativistic fluid (photons, neutrinos) the energy density and the pressure are given by 
 \begin{align}
  \rho_{r}&=4\sigma T_{r}^{4},\\
   p_{r}&=\dfrac{\rho_{r}}{3},
 \end{align}
where $\sigma$ is the Stefan-Boltzmann constant. From Eq.~(\ref{eq:QthEq}), the interaction  term needed to maintain  thermal equilibrium follows immediately:
  \begin{equation}
      \tilde{Q}_{r}=\left[\dfrac{4HR-4\dot{R}}{R}\right]\rho_{r}.
     \label{eq:Qr}
  \end{equation}
We note that by replacing the variables associated with dark energy in Eq.~(\ref{eq:Qwde}) by those associated with radiation, we get the same expression. This is not surprising since to obtain Eq.~(\ref{eq:Qwde}),  we considered a fluid with a constant EoS parameter and whose energy density depends only on the temperature, as  is the case  for  radiation ($\rho_r=4\sigma T_r^4$, $w_r=1/3$). More generally, all the results of section~\ref{sect:wDE} hold for any fluid fulfilling these two conditions, which excludes however non-relativistic matter. In particular, Eq.~(\ref{eq:rwde}) becomes for radiation
 \begin{equation}
      \rho_{r}=\rho_{r_0}\left(\dfrac{R}{R_{0}}\right)^{-4}.
    \label{eq:rr}
  \end{equation}

For a non-relativistic fluid, such as dark matter or baryonic matter, the energy density and the pressure are given by
 \begin{align}
  \rho_{m}&=n_mm+\frac{3}{2}n_m T_{m},\label{eq:rm}\\
  p_{m}&=n_m T_{m}\label{eq:pm},
 \end{align}
where $n_m\equiv N_m/V$ is the particle number density. Here we consider a single particle species of mass $m$, but the generalization to many species is straightforward. Inserting  Eqs.~(\ref{eq:rm}) and (\ref{eq:pm}) into Eq.~(\ref{eq:QthEq}) leads to
 \begin{equation}
    \tilde{Q}_m=\left[\dfrac{3w_m HR-\frac{3}{2}w_m\dot{R}}{R}\right]\rho_m,
  \label{eq:Qm}
  \end{equation}
where the EoS parameter is given by
  \begin{equation}
      w_m\equiv\dfrac{p_m}{\rho_m}=\dfrac{T_m}{m+\frac{3}{2}T_m}.
  \end{equation}
Assuming that the rest-energy of the fluid is much larger than its kinetic energy (${m\gg T_{m}}$), the EoS parameter may be approximated by $w_m\approx T_m/m$. Since ${w_m\ll1}$,  cold matter is usually considered to be pressureless ($w_m=0$). However, we cannot use this approximation here since that would imply, according to Eq.~(\ref{eq:Qm}), that $\tilde{Q}_m=0$. Using Eq.~(\ref{eq:Thor}), the  EoS parameter may be written more conveniently as a function of the horizon radius
  \begin{equation}
      w_m=w_{m_0}\dfrac{R_0}{R}, 
  \end{equation}
where $w_{m_0}\equiv b/(2\pi m R_0)$. Inserting the interaction term $\tilde{Q}_m$ into the continuity equation (\ref{eq:cont}) and solving it yields
 \begin{equation}
    \rho_m=\rho_{m_0}a^{-3}\exp{\left[\frac{3}{2}w_{m_0}\left(\dfrac{R_0-R}{R}\right)\right]}.
  \label{eq:rhom}
  \end{equation}

Now we must check whether the interaction terms found are consistent with the energy conservation condition $\sum Q_i=\sum \tilde{Q}_{i_{\rm eq}}+\sum Q_{i_{\rm neq}}=0$. The summation indices ${i_{\rm eq}}$ and ${i_{\rm neq}}$ refer respectively to the fluids that are in thermal equilibrium with the horizon, and to those that are not. In the case where at least one of the interacting fluid is not in equilibrium, we can set $\sum Q_{i_{\rm neq}}=-\sum\tilde{Q}_{i_{\rm eq}}$ in order to fulfill the energy conservation condition. However, when all the interacting fluids are assumed to be in thermal equilibrium we must have $\sum \tilde{Q}_{i_{\rm eq}}=0$, from which we get the following expression for the Hubble rate
 \begin{equation}
    H=\left[\dfrac{\sum\beta_{i_{\rm eq}}\rho_{i_{\rm eq}}}{3\sum(1+w_{i_{\rm eq}}-\delta_{{i_{\rm eq}}}^{m})\rho_{i_{\rm eq}}}\right]\dfrac{\dot{R}}{R},
  \label{eq:Hgen}
  \end{equation}
where $\beta_i=2, (1+w_{wde})^{-1}, 4$ and $\frac{3}{2}w_{m}$ respectively for HDE, $w$DE, radiation and cold matter.  The value of  $\delta^m_i$ is 1 when $i=m$ and 0 otherwise. The energy density of the fluids in thermal equilibrium (Eqs.~(\ref{eq:hde}), (\ref{eq:rwde}), (\ref{eq:rr}) and (\ref{eq:rhom})) depends only  on the horizon radius $R$ and on the scale factor $a$ (for cold matter), hence Eq.~(\ref{eq:Hgen}) can be integrated (at least numerically) in order to find the relationship between these two variables. However, the function $R(a)$ thus obtained does not necessarily coincide with one of the three radii ($R_H$, $R_A$, $R_E$) considered in section~\ref{sect:cht}. 

To illustrate the previous statement, we will consider the case  where  $w$DE and  radiation are  in thermal equilibrium and are the only two interacting fluids.  This example is among the simpler to consider  because  Eq.~(\ref{eq:Hgen}), which becomes
 \begin{equation}
    H=\left[\dfrac{4\rho_r+(1+w_{wde}^{-1})\rho_{wde}}{4\rho_r+3(1+w_{wde})\rho_{wde}}\right]\dfrac{\dot{R}}{R},
  \label{eq:Hwder0}
 \end{equation}
can be integrated analytically. Inserting the expressions found for $\rho_r$ and $\rho_{wde}$ (Eqs.~(\ref{eq:rwde}) and  (\ref{eq:rr})) gives
 \begin{equation}
    H=\left[\dfrac{4r_{r_0} + (1+w_{wde}^{-1})\tilde{R}^{3-w_{wde}^{-1}}  }{4r_{r_0} + 3(1+w_{wde})\tilde{R}^{3-w_{wde}^{-1}} }\right]\dfrac{\dot{\tilde{R}}}{\tilde{R}}.
  \label{eq:Hwder}
  \end{equation}
Here, we have introduced the dimensionless radius $\tilde{R}=R/R_0$ and the radiation to dark energy density ratio at $t_0$ ($r_{r_0}\equiv\rho_{r_0}/\rho_{wde_{0}}$) . Integration of Eq.~(\ref{eq:Hwder})  yields
 \begin{equation}
    a=\left[\dfrac{4r_{r_0}+3(1+w_{wde})}{4r_{r_0}+3(1+w_{wde})\tilde{R}^{3-w_{wde}^{-1}}}\right]^{\frac{1}{3}}\tilde{R}.
  \label{eq:awder0}
  \end{equation}
By differentiating this equation, we find that the scale factor reaches a maximum value $a_{max}$ at
 \begin{equation}
    \tilde{R}_{a_{max}}=\left(-\dfrac{4r_{r_0}w_{wde}}{1+w_{wde}}\right)^\frac{1}{3-w_{wde}^{-1}}.
  \label{eq:Ramax}
  \end{equation}

Consistently, the expression for the Hubble rate given by Eq.~(\ref{eq:Hwder0}) is zero at $\tilde{R}=\tilde{R}_{a_{max}}$.  The expression for the Hubble rate given by the first Friedmann equation (Eq.~(\ref{eq:fried1})) must also be zero at this point. This condition reduces by one the number of free parameters in the model. For instance, we can express the value of the energy density of the spatial curvature as 
 \begin{equation}
   {\rho_{k_0}=-\left.\sum_{i\neq k}\rho_i a^2\right|_{\tilde{R}=\tilde{R}_{a_{max}}}},
  \end{equation}
where the energy density of the non-interacting fluids ($i\neq wde, r$) is given by Eq.~(\ref{eq:rho}). Not surprisingly for a cosmic scenario involving  recollape, we find that the spatial curvature is positive ($\rho_{k_0}<0$). The value of the remaining parameters can be chosen freely (provided that $\sum_{i\neq k}\rho_{i_0}+\rho_{k_0}\geq0$, in order to have $H_0\in\mathbb{R}$) and leads to a self-consistent cosmology where the radiation and $w$DE and are in thermal equilibrium with a cosmological horizon whose radius is implicitly defined in Eq.~(\ref{eq:awder0}). Now, we want to verify wether this radius  coincides either with the Hubble radius, the apparent radius or the event horizon radius. By solving the equation $\tilde{R}=\tilde{R}_{H}(\tilde{R})=1/|H(\tilde{R})|$ for the constant $\rho_{r_0}$, we get
 \begin{equation}
   \rho_{r_0}=-\left[
   \begin{split}
   &\sum_{i\neq r,wde}\rho_{i_0}a(\tilde{R})^{-3(1+w_i)}\\
     &+\rho_{wde_0}\tilde{R}^{-(1+w_{wde}^{-1})}-3M_p^2\tilde{R}^{-2}
  \end{split}
   \right]\tilde{R}^4.
  \end{equation}
Solving  $\tilde{R}=\tilde{R}_{A}(\tilde{R})$ for $\rho_{r_0}$ leads to the same expression, except that now, the spatial curvature is  excluded from the summation ($i\neq k,r,wde$). In both cases, we obtain an expression for the constant $\rho_{r_0}$ which is actually a function of $\tilde{R}$. This inconsistency shows that $\tilde{R}\neq\tilde{R}_{H}$ and $\tilde{R}\neq\tilde{R}_{A}$. For the event horizon radius, we cannot directly compare  $\tilde{R}_E$ to $\tilde{R}$  by reason of the integral involved in the definition of this radius. However, we can compare its time derivative, which is
 \begin{equation}
   \dot{\tilde{R}}_E=\dfrac{\dot{R}_E}{R_{E_0}}=H\tilde{R}_E-R_{E_0}^{-1},
  \end{equation}
to the expression for $\dot{\tilde{R}}$ obtained from Eq.~(\ref{eq:Hwder}). Solving $\dot{\tilde{R}}=\dot{\tilde{R}}_E$ for $R_{E_0}$ and replacing $\tilde{R}_E$ by $\tilde{R}$ yields
 \begin{equation}
 \begin{split}
   &R_{E_0}=\left[\dfrac{\sqrt{3}M_p}{1-2w_{wde}-3w_{wde}^2}\right]\times\\
&\left[\dfrac{4r_{0}w_{wde}\tilde{R}^{w_{wde}^{-1}-4}+(1+w_{wde})\tilde{R}^{-1}}{\sqrt{\sum\limits_{i\neq r,wde}\rho_{i_0}a(\tilde{R})^{-3(1+w_i)}+\rho_{r_0}\tilde{R}^{-4}+\rho_{wde_0}\tilde{R}^{-(1+w_{wde}^{-1})}}}\right].
 \end{split}
 \end{equation}
Once again, we obtain an inconsistent equation where a constant is equal to a function of $\tilde{R}$, showing that $\tilde{R}\neq{\tilde{R}}_E$. Here we have shown that none of the three radius definitions considered in section~\ref{sect:cht} could lead to thermal equilibrium between  the cosmological horizon, radiation and $w$DE if the other fluids are not interacting.  More generally, when a different combination of fluids is considered, we should proceed similarly to this example and verify whether the radius obtained from Eq.~(\ref{eq:Hgen}) is meaningful or not.

\section{IV. Summary}
\label{sect:Conclusion}

When the thermodynamical properties of dark energy are studied, the hypothesis of (late time) thermal equilibrium between the cosmological horizon and the dark energy fluid is frequently assumed \cite{ThEq0,ThEq1,ThEq2,ThEq3,ThEq4,ThEq5,ThEq6,ThEq7,ThEq8,FF01,FF02,FF03,N02} and, in some cases, even extended to other cosmological fluids \cite{ThEq5,ThEq6,ThEq7,ThEq8,FF01,FF02,FF03}. The aim of this paper was to find the restriction imposed by this hypothesis on the energy transfer rate ($Q_i$) between the fluids.  

A first difficulty occurs in defining the  temperature of the horizon. In a dynamical spacetime, such as the FRW spacetime, there is no consensus for which horizon (if any)  should emit  Hawking radiation and, for a given choice, what should be the temperature associated with this radiation. In order to recover different expressions used in the literature, we have considered a temperature of the form $T_h=b/2\pi R$, where $R$ could stand for the Hubble radius ($R_H$) \cite{ThEq4,FF03,Tprop03}, for the apparent radius ($R_A$) \cite{ThEq1,ThEq3,ThEq4,ThEq7,FF03} or for the event horizon radius $(R_E)$ \cite{ThEq0,ThEq2,ThEq5}. 

A second difficulty is the unknown nature of dark energy. We  considered a generic fluid to find the interaction term required to maintain  thermal equilibrium (Eq.~(\ref{eq:QthEq})), but to go further in  our analysis, we specialized to two specific types of dark energy, namely holographic dark energy (HDE) and  dark energy with a constant EoS parameter ($w$DE). For HDE, in thermal equilibrium, the energy density is only a function of the temperature and we  assumed that was also the case for its pressure. We made the same assumption for  $w$DE dark energy. This leads to interaction terms given by Eq.~(\ref{eq:Qhde}) for HDE and Eq.~(\ref{eq:Qwde}) for $w$DE. These results illustrate that, in general, if we assume thermal equilibrium between the dark energy and a horizon of radius $R$, we cannot choose the interaction term $Q_i$ freely. Conversely, if we impose a specific choice for the interaction term, the radius $R$ will be determined by inverting these equations, which will not necessarily correspond to a physically meaningful  horizon. 

Finally, we  found the interaction terms for which radiation (Eq.~(\ref{eq:Qr})) and cold matter (Eq.~(\ref{eq:Qm})) are in thermal equilibrium with the horizon. Since the ensemble of the interaction terms must satisfy $\sum_i Q_i=0$, it is non-trivial to propose a cosmological model for which all the interacting fluids are in thermal equilibrium with the horizon. Indeed, in this case, the horizon radius will be determined by Eq.~(\ref{eq:Hgen})  and will not necessarily be physically meaningful.

\section{ACKNOWLEDGMENTS}
The author want to thank James Cline for his comments on this manuscript. This work was partly supported by the Fonds de recherche du Qu\'ebec - Nature et technologies (FQRNT) through the doctoral research scholarships programme. 


\end{document}